\documentclass[conference]{IEEEtran}
\IEEEoverridecommandlockouts
\usepackage{cite}
\usepackage{amsmath,amssymb,amsfonts}
\usepackage{algorithmic}
\usepackage{graphicx}
\usepackage{textcomp}
\usepackage{xcolor}
\usepackage{float}
\usepackage{booktabs}
\usepackage{url}

\def\BibTeX{{\rm B\kern-.05em{\sc i\kern-.025em b}\kern-.08em
    T\kern-.1667em\lower.7ex\hbox{E}\kern-.125emX}}

\begin{document}

\title{
Mixing-Specific Data Augmentation Techniques for Improved Blind Violin/Piano Source Separation}


\author{\IEEEauthorblockN{Ching-Yu Chiu}
\IEEEauthorblockA{\textit{Graduate Program of Multimedia Systems and Intelligent Computing} \\
\textit{National Cheng Kung University and Academia Sinica}, Taiwan\\
sunnycyc@citi.sinica.edu.tw}
\and
\IEEEauthorblockN{Wen-Yi Hsiao}
\IEEEauthorblockA{\textit{Yating Music Team} \\
\textit{Taiwan AI Labs}, Taiwan\\
wayne391@ailabs.tw}
\and
\IEEEauthorblockN{Yin-Cheng Yeh}
\IEEEauthorblockA{\textit{Yating Music Team} \\
\textit{Taiwan AI Labs}, Taiwan\\
yyeh@ailabs.tw}
\and
\IEEEauthorblockN{Yi-Hsuan Yang}
\IEEEauthorblockA{\textit{Research Center for IT Innovation, Academia Sinica}, Taiwan \\ 
yang@citi.sinica.edu.tw}
\and
\IEEEauthorblockN{Alvin Wen-Yu Su}
\IEEEauthorblockA{\textit{Dept. CSIE, National Cheng Kung University}, Taiwan \\
alvinsu@mail.ncku.edu.tw}
}

\maketitle

\begin{abstract}
Blind music source separation has been a popular and active subject of research in both the music information retrieval and signal processing communities. To counter the lack of available multi-track data for supervised model training, a data augmentation method that creates artificial mixtures by combining tracks from different songs has been shown useful in recent works. Following this light,  we examine further in this paper extended data augmentation methods that consider  more sophisticated mixing settings employed in the modern music production routine, the relationship between the tracks to be combined, and factors of silence. As a case study, we consider the separation of violin and piano tracks in a violin piano ensemble, evaluating the performance in terms of common metrics, namely SDR, SIR, and SAR. In addition to examining the effectiveness of these new data augmentation methods,  we also study the influence of the amount of training data.  Our evaluation shows that the proposed mixing-specific data augmentation methods can help improve the performance of a deep learning-based model for source separation, especially in the case of small training data. 
\end{abstract}


\section{Introduction}
\label{sec:intro}
Music source separation, i.e., separating the sources (instruments) involved in an audio recording, has been a major research topic in the signal processing community, partly due to its wide downstream applications in music upmixing and remixing, karaoke, DJ-related applications, and as a pre-processing tool for other problems \cite{vincent06taslp,yang13ismir,jao16taslp,tachibana2016real,Uhlich2017,jansson2017singing,stoller2018adversarial,Rafii2018,Liu2018,Liu2019,Brocal19,leeCL19,Stoter2019,Hennequin2019,Manilow2020,hsieh20icassp}.  
Its technical difficulty has also been well acknowledged. For example, 
the estimation of the number of sources involved in a song remains challenging \cite{luo17icassp}.  
Even when the number of sources is known or given beforehand, in supervised training we need the \emph{multi-track recordings} that provide the ground truth single-instrument tracks (a.k.a., `stems') that compose a song.  Such multi-track recordings are rarely publicly available due to copyright issues \cite{Bittner2014,Bittner2016a}. 
What are typically available instead are the \emph{mixed} versions of the songs, where the multiple tracks have been mixed and combined into a monaural (i.e., one-channel) or stereo (two-channel) recording. The lack of available data with ground truth  stems not only limits the development of data-driven methods, but also hinders systematic evaluation of new methods proposed for the task.   

Over the past decades, the main methods for tackling this task can be roughly classified into two categories: model-based methods and data-centered methods. As discussed in a recent review paper \cite{Rafii2018}, the performance of model-based methods could change dramatically when their core assumptions are not met. On the other hand, data-centered methods rely heavily on the availability of professionally produced or recorded multi-track data, which is hard to come by due to copyright issues.

\begin{figure}
\centerline{\includegraphics[width = 1.0\columnwidth]{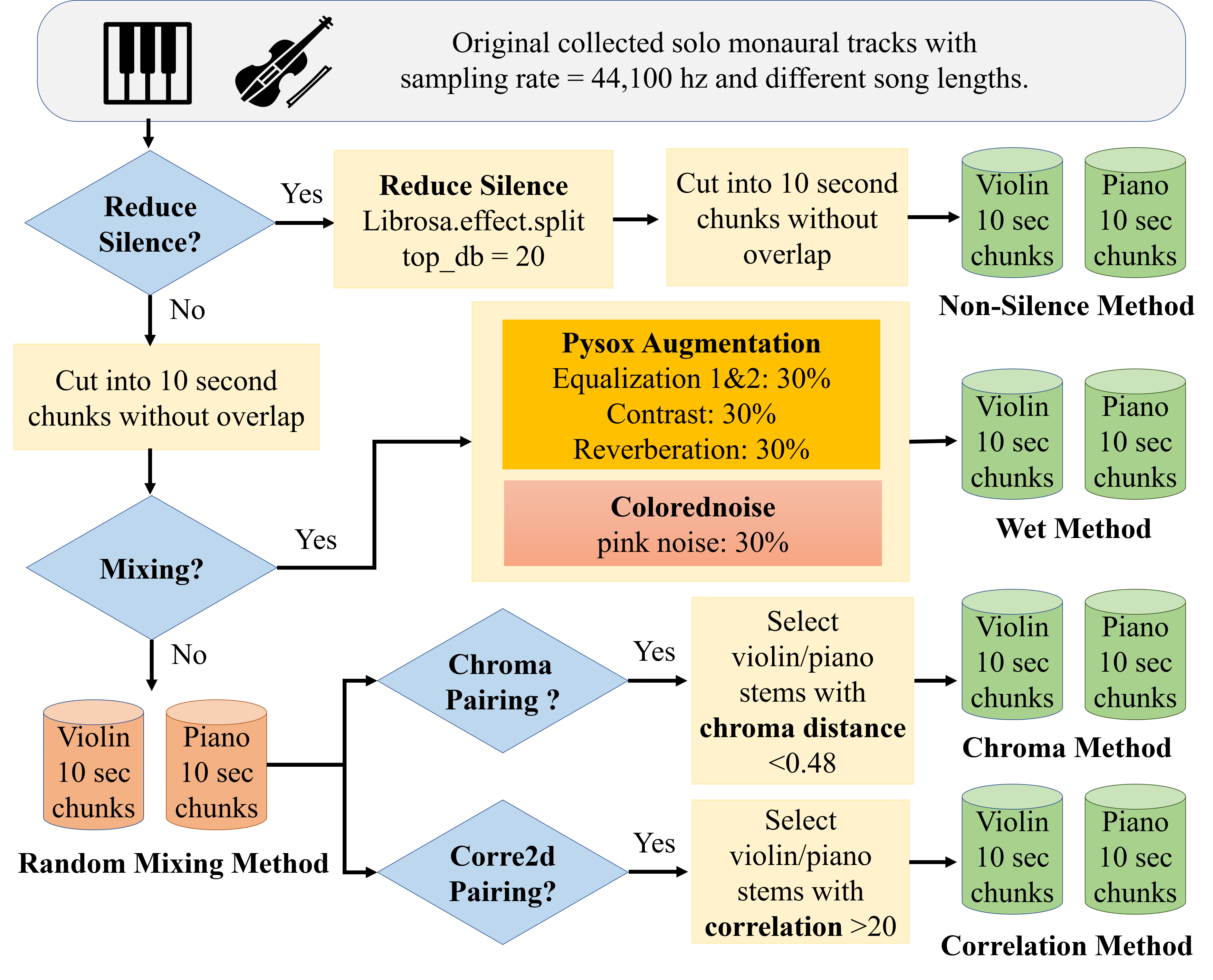}}
\caption{ Flowchart of the baseline random mixing (bottom-left) and the proposed four augmentation methods (on the right). Given  the collected violin solos and piano solos (in gray block), three types (Silence, Mixing, and Pairing) of augmentations (in blue blocks) can be applied to produce four groups of augmented violin/piano chunks (in green buckets). If none of them is applied, we use the violin/piano chunks (in orange bucket) for the baseline random mixing. Best viewed in color.  
}
\label{fig:methods}
\end{figure}

As some medium-scale multi-track datasets have been released in the past few years, the development of data-driven models grows fast. A data-driven source separation model is typically a supervised model which is trained by taking an audio mixture (a monaural or stereo recording) as the input, and aiming to recover the tracks (i.e., multiple monaural or stereo recordings) that compose the mixture.  In doing so, there have been two main approaches in the literature. In the first approach, which we refer to as the \textbf{no data augmentation} approach, the tracks that compose an input mixture are originally from the same song. In other words, when we have $N_0$ multi-track songs, we would have exactly $N_0$ input/output pairs for modeling training. In the second approach, or \textbf{random-mixing data augmentation} \cite{Stoter2019,Uhlich2017},  tracks from different songs are randomly combined to create audio mixtures, leading to artificial input/output pairs. In such a case, we can have $N \gg N_0$ input/output pairs.  The downside of this approach is these input mixtures are not realistic sound mixtures in terms of the tonic, harmonic and rhythmic relations among the tracks that compose the mixtures. But, for the purpose of training data-driven models for source separation, the benefit of the resulting great increase in the number of training data seems to outweigh this potential concern, as demonstrated in the literature \cite{Stoter2019,Uhlich2017}. 

While there are some other data augmentation methods such as adding noise or randomly dropout  \cite{erdogan2018, Schluter2015}, the aforementioned \emph{random-mixing data augmentation}, albeit simple, has been shown particularly successful \cite{Stoter2019}.  However, when mixing two tracks, there are actually multiple aspects to consider \cite{Man2017TowardsAB}, suggesting room for the development of more advanced mixing-specific data augmentation methods for source separation. To our best knowledge, this has not yet been investigated in the literature. It is therefore our goal to develop, and to empirically evaluate the effectiveness of, new data augmentation methods that stem from the random-mixing approach.  We consider in total three types of mixing-specific data augmentation methods for source separation, as conceptually visualized in Figure \ref{fig:methods} and detailed in Section \ref{sec:method}. 

Besides, current available data is still not enough for many common musical instruments, such as the violin. To investigate the benefit of data augmentation for music source separation in general, we consider in this paper the separation of violin and piano tracks in a violin piano ensemble, a task that has rarely been considered in the literature.  Moreover, we consider the case when \emph{no multi-track recordings of violin piano ensembles} are available for model training, but instead \emph{only a collection of piano solos and violin solos}. In such a case, mixing-based data augmentation becomes a major viable approach.  

In sum, the main contribution of this study is to propose a series of data augmentation/selection approaches that enable \emph{non-paired violin/piano solo stems} to approximate features of realistic paired stems, which in turn  facilitate the training of deep learning-based source separation models.

For reproducibility, we share the code, pre-trained models, the audio files of the ground truth and  separated stems (by the `Wet' model; see Section \ref{sec:method}) of the test data publicly at \url{https://github.com/SunnyCYC/aug4mss} and \url{https://sunnycyc.github.io/aug4mss_demo/}.

Below, 
we review  related work and the adopted network architecture for  source separation in Section \ref{sec:related}. Section \ref{sec:db} describes the training and test data employed in our implementation. Section \ref{sec:method} presents the proposed data augmentation methods, while Section \ref{sec:exp} talks about the evaluation results. Finally, we conclude the paper in Section \ref{sec:final}.

\begin{figure}
\centerline{\includegraphics[width = 0.84\columnwidth]{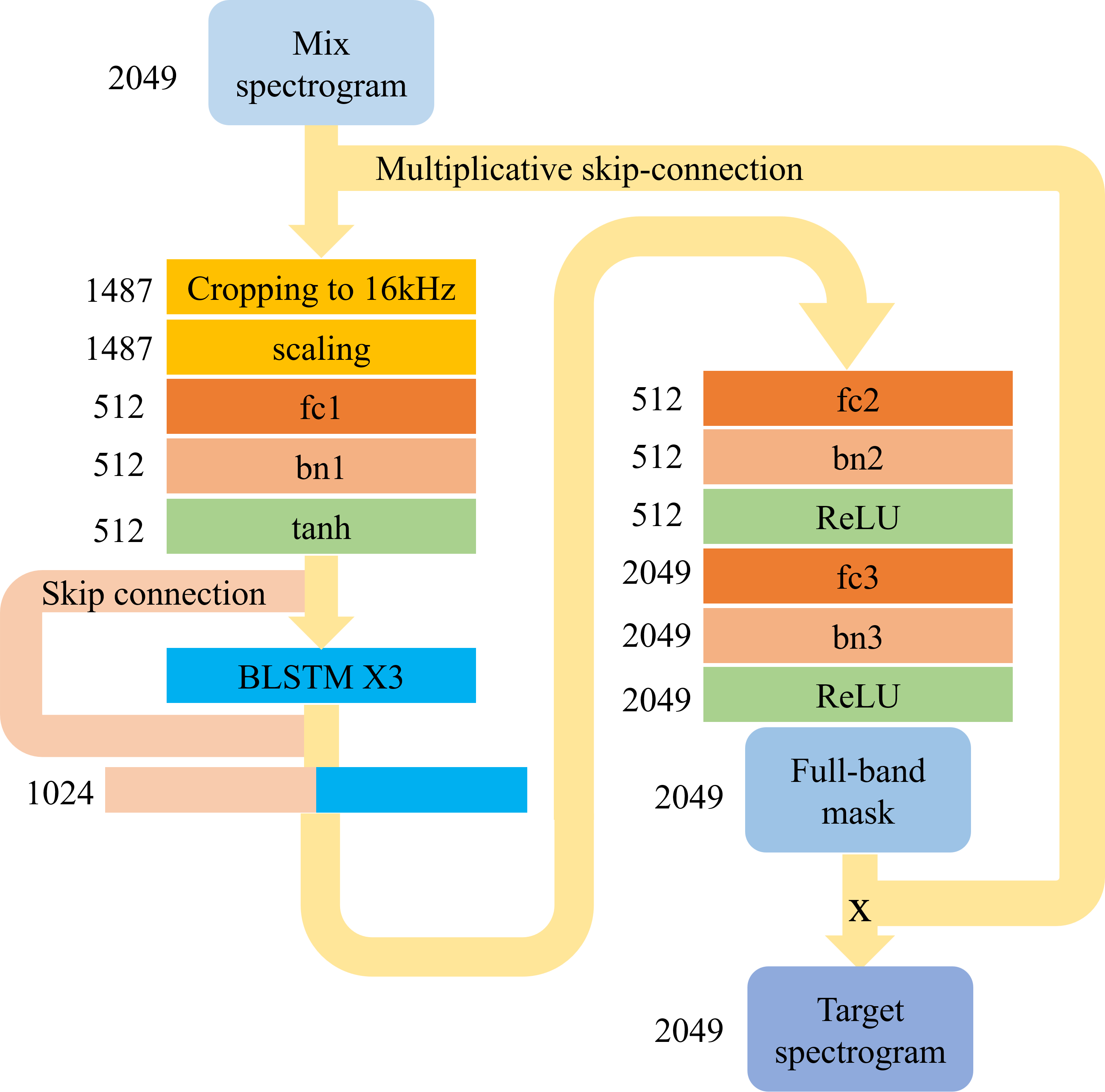}}
\caption{The network architecture of Open-Unmix  \cite{Stoter2019}. Taking an input mix spectrogram, the model produces the target spectrogram by multiplying the input with a full-band mask. The model is based on a three-layer bidirectional LSTM capable of taking input of arbitrary lengths. The number on the left side of each component indicates the frequency bin size of the output. The model crops and only processes the spectrogram under 16kHz, and reconstructs intermediate product to full band by the last fully connected layer `fc3.'} 
\label{fig:openunmix}
\end{figure}

\section{Background}
\label{sec:related}

\subsection{Related works}
Data augmentation for improving the performance of deep neural networks has been an important topic in the field of music information retrieval (MIR) in recent years. Schluter and Grill \cite{Schluter2015} were one of the earliest to systematically explore the utility of music data augmentation for singing voice detection with neural networks. They found pitch shifting combined with time stretching and random frequency filtering to be quite helpful in reducing the classification error.
Uhlich et al. \cite{Uhlich2017} proposed two neural network architectures capable of yielding state-of-the-art results for music source separation at that time, and further boosted the performance through data augmentation and network blending. Hawthorne et al. \cite{Hawthorne2019} experiments with the use of mixing techniques such as equalization, contrast, and reverberation in an attempt to make their automatic piano transcription model more robust to different recording environments and piano qualities. 
To our knowledge, such mixing techniques have not been employed in existing work on source separation.

New neural network architectures for blind music source separation have been continuously proposed as well. For example, Manilow et al. \cite{Manilow2020} intellectually utilized the inherent synergy between transcription and source separation to improve both tasks using a multi-task learning architecture. Liu and Yang \cite{Liu2019} combined dilated convolution with modified gated recurrent units (GRU) to extend the receptive field of each dilated GRU unit, enabling their model to perform better and faster than state-of-the-art models for separating vocals and accompaniment.

\subsection{Model Architecture}
As our focus is on the data augmentation techniques, we adopt an existing blind source separation framework called Open-Unmix \cite{Stoter2019} as the backbone architecture in our work. Open-Unmix is a hybrid convolutional-recurrent architecture (see Figure \ref{fig:openunmix}) that is open source.
It takes a fixed-length chunk of the Short-time Fourier transform (STFT) spectrogram of the mixture as the input and aims to get the corresponding separated spectrogram of one of the sources at the output. Although the model is trained on fixed-length chunks, at testing time it can be applied to spectrograms of arbitrary length. The parameters of the network is learned by minimizing the difference between the spectrograms of the ground truth and the separation result, calculated in terms of mean square error.

The original Open-Unmix model does not deal with the separation of piano and violin tracks \cite{Stoter2019}, but it is easy to use their code base to train the model on our piano and violin data. In comparison, the other famous open-source separation model,  called Spleeter \cite{Hennequin2019}, has a built-in function to isolate out the piano track from an input mixture, but it does not provide the script for retraining the model to isolate out the violin track. In consequence, we consider the pre-trained Spleeter model released by the authors as the baseline method for performance comparison, instead of the backbone architecture of our model. 

\section{Datasets}
\label{sec:db}

\subsection{Violin/piano Solo as Training Stems }

We collected six hours of classical violin solo recordings and six hours of pop piano solo recordings from the Internet as our training and validation data. For each instrument, five hours of data is allocated for training and the remaining one hour for validation. All songs are divided into 10-second chunks. During training and validation, one chunk will be randomly selected from each instrument for mixing and serving as the input data. 

\subsection{MedleyDB as Evaluation Data}
To evaluate the performance of our method on real data, we select multi-track songs that contain realistic violin and piano stems from the MedleyDB \cite{Bittner2014,Bittner2016a} for evaluation. There are only 16 such songs, highlighting the difficulty of collecting multi-track recordings for model training. Moreover, after listening to these songs one-by-one, we discard 10 of them, as there are severe leakage issues and accordingly the  piano/violin stems are not purely piano/violin.  We also note that most of the remaining six songs contain more than one violin stems. For such songs, we consider the combination of the piano track and \emph{one} of the violin tracks as the input mixture, thereby creating multiple partially realistic mixtures from the same song. All the other instruments from these songs (e.g., cello) are also excluded.   As a result, we have 16 violin piano ensembles in total for evaluation.

We note that, due to copyright restrictions, we are unfortunately not able to share the audio files of the training data. Researchers interested in violin piano separation would have to collect the training data on their own. 
However, as mentioned by the end of Section \ref{sec:intro}, we make public the aforementioned 16 violin piano ensembles so that people can evaluate their models on the same test set.

\begin{table}
\caption{Three types of mixing-related data augmentation methods we propose here, and the corresponding parameter settings}
\centering
\begin{tabular}{ll|ll}
\toprule
\textbf{Mixing} &   Description                & Scale     & Range \\
\midrule
       &   Contrast   & Amount & 1$-$70  \\
       &   Equalizer 1             & Frequency      & 32$-$4,096  \\
       &   Equalizer 2             & Frequency      & 32$-$4,096  \\
       &   Reverb             & Reverb      & 1$-$70  \\
       &   Pink noise  & Volume & 0.01$-$0.04   \\
\toprule
\textbf{Paring} &   Description                & Threshold   & Notes\\
\midrule
       &   Chroma distance            & 0.48        & Mean within songs: 0.45\\
       &                              &             & Mean across songs: 0.51\\
       &   Correlation                & 20          & Within songs: 2$-$30\\
       &                              &             & Across songs: 0$-$10\\
\toprule
\textbf{Silence}&   Description                & Threshold   & Notes\\
\midrule
       &   Reduce silence             & 20          & top db\\
\bottomrule
\end{tabular}
\label{tab:model}
\end{table}

\section{Proposed Augmentation/Selection Methods}
\label{sec:method}
As shown in Figure \ref{fig:methods} and Table \ref{tab:model}, we consider three types of augmentation methods here. 
The \textbf{mixing type} augmentation includes factors related to the common mixing process, such as the use of equalization, contrast, reverb, and the addition of pink noise. The \textbf{pairing type} methods are designed based on consideration for the tonic, harmonic and rhythmic relations between the real paired stems. Lastly, the \textbf{silence type} method is designed for eliminating the difference of silence duration in violin/piano stems to avoid potential imbalance of the training data. In what follows, we refer to the unprocessed original stems as the \emph{original stems}.

\subsection{Wet Stems}
Following \cite{Hawthorne2019}, we apply the common approaches in the mixing process and also the addition of pink noise as our first augmentation method. The pink noise is employed to simulate the background noise seen in real recordings. The augmentation parameters are shown in Table \ref{tab:model}. For each original stem, we set a 30\% probability to apply a specific process (e.g. equalization). The pink noise is applied using \texttt{colorednoise}, and the others are applied using a python package called \texttt{pysox} \cite{Bittner2016}. In what follows, we also refer to a model trained with this data augmentation method (i.e., the `Wet' method) as the \textbf{Wet model}.

\subsection{Chroma Distance-based Pairing}
Based on general principles of music theory or psychoacoustics, the paired stems in music are usually highly coherent in pitch, or are in similar keys. Therefore, instead of randomly selecting stems for combination, this augmentation method only picks the stems that have short chroma distance with one another for combination, in a hope that the resulting mixture would be  more similar to real ensembles. Specifically, we average the chromagram, a representation of the time-varying intensities of the twelve different pitch classes, to derive a 12-dimensional chroma feature for each stem, and then calculate the Euclidean distance between all violin/piano stems. We then need a threshold for selecting qualified training violin/piano stems. In doing so, we rely on statistics calculated from the MedleyDB test songs. As shown in Table \ref{tab:model}, the mean chroma distance between the violin/piano stems from the same song of MedleyDB is 0.45, shorter than the mean distance of violin/piano stems from different songs of MedleyDB. We therefore set the threshold to 0.48. Only stems with chroma distance lower than this threshold will be selected and mixed as our training data.

\subsection{Correlation-based Pairing}
Another pairing method is to consider whether the piano stem and violin stem are \emph{active} (i.e., non-silent) at the same time; namely, whether they co-occur.   To implement this, we calculate the absolute value of the 2-dimensional cross correlation between the waveform magnitude (using \texttt{scipy.signal.correlate2d}) of all the violin/piano stems and set a threshold for selecting the stems to be combined for training. As shown in Table \ref{tab:model}, the cross correlation values for true paired stems in MedleyDB test songs range from 2--30. We therefore empirically set the threshold to 20.

\subsection{Silence Removal before Mixing}
For the case of violin piano ensemble, we empirically observe sometimes the activity of the violin part would be too sparse, making the two sounds unbalanced in the mixture. We consider it worth investigating whether this would influence the performance of the resulting separation model. Therefore, we apply \texttt{librosa.effects.split} \cite{McFee2015} with
`top db’ $=$ 20 to remove the silence part in the training/validation data, before dividing them into 10-second chunks for mixing. 
We have also experimented with other values of `top db’ and found 20 works better.

\section{Experiments}
\label{sec:exp}
\subsection{Experiment Settings \& Evaluation Metrics} 
As discussed in Section \ref{sec:intro}, there are  many common instruments that suffer from the lack of multi-track data. 
For instruments less common, such as 
erhu and suona, 
even the number of solo recordings is  limited.
Therefore, 
we study the effectiveness of the proposed augmentation methods in two  scenarios: a \textbf{data-limited} one, and a \textbf{data-rich} one. 

Specifically, for each instrument and each augmentation method, we train (from scratch) a separation model using the Open-Unmix architecture and the corresponding processed stems. In each training epoch, the model will randomly select $N$ pairs of stems from the pool of processed stems to mix as the training data. To simulate the data-limited case, we set $N=$~250 and adopt only 16 minutes of the training data for each instrument. For the data-rich case, we use the full data for training and set $N=$~2000.
After each training epoch, the model will also select 100 pairs of stems from the validation pool of processed stems to mix as validation data. The validation loss is adopted by an early stop mechanism and a learning rate scheduler to fine-tune the whole training process.
For all experiments, an early stop patience, 140 epochs, is adopted, and the learning rate is set to 0.001. All the training and testing songs are \textbf{monaural} tracks with a 44,100 hz sampling rate. The window size and hop length of STFT are 4,096 and 1,024. The predicted spectrograms are  converted back to time-domain waveforms using the Griffin-Lim algorithm 
for phase estimation before inverse STFT. 

For evaluation metrics, we adopt the 
signal-to-distortion ratio (SDR), signal-to-interference ratio (SIR), and signal-to-artifacts ration (SAR), implemented in the \texttt{BSS\_eval} toolkit \cite{vincent06taslp}.
Following the convention in SiSEC (\url{https://sisec.inria.fr/}), we report the median values over the testing songs.

We consider the following two methods as the baseline methods. First, we use the Open-Unmix model to train the separation model using the existing random-mixing data augmentation method; we refer to this method as the \textbf{Random} method. Second, we use the official pretrained 5-stem model of \textbf{Spleeter} \cite{Hennequin2019}, 
the current state-of-the-art for singing voice separation.
Specifically, we use the default `mask'-based implementation.  Given an input mixture, it generates as the output separated stems of piano, vocal, drum, bass, and others. From the result of Spleeter, we sum all but the piano stem as the separated violin stem. 

Training an Open-Unmix based network using any of the proposed data augmentation method takes around 15 hours on an NVIDIA GeForce GTX 1080 GPU. At testing time, it takes $\sim$3 mins to complete the separation (for both the piano and violin) of an 18-min violin piano ensemble, on the same GPU.


\subsection{Results \& Discussion}
The result of the data-limited case is shown in Table \ref{tab:result1}.  The following observations can be made:
First, for piano, all the proposed augmented models outperform the two baselines in SDR and SAR. In particular, the improvement made by `Correlation' method is more than 2 dB for both metrics.
Second, for violin, except for the `Correlation' method that still gains improvement in SDR and SIR, the help of other augmentation methods becomes less obvious. The performance drop of `Wet' method for violin indicates potential risk for the randomly applied mixing type method to distort the data under data-limited scenario.
Finally, all the Open-Unmix based models we train greatly outperform the pre-trained, violin-agnostic Spleeter model in almost all the metrics for both instruments, except for the piano in terms of SIR. 
    This is not surprising, as the Spleeter model has not been specifically trained on violin data. The failure of recognizing the violin also hurts its performance on the piano. 


The result of the data-rich case are shown in Table \ref{tab:result2}.
From all the metrics we can see that the help of augmentations in both instruments becomes less obvious. For the piano, the baseline `Random’ method seems strong enough. For the violin, the baseline `Random’ method is only inferior to the proposed method with a small margin.  This implies that, when the training data is big enough, the diversity of the data would also increase and overshadow the help of sophisticated augmentation methods.

\begin{table}
\caption{The median values of the evaluation metrics (in db) over the 16 MedleyDB songs for the data-limited scenario} 
\centering
\begin{tabular}{l|lrrr}
\toprule
\textbf{Piano} & Method &  SDR & SIR & SAR \\
\midrule
               & Spleeter (pretrained)  \cite{Hennequin2019}   & 5.20  & \textbf{17.35} & 2.30\\
\hline
                & Random-mixing   & 7.43 & 14.38 & 10.52\\
\hline
               & Wet    &8.48 & 13.80 & 11.21\\
               & Chroma & 7.47 & \ \ 13.09 & 10.91\\ 
               & Correlation & \textbf{9.66} & 13.89 & \textbf{13.80}\\
               & NonSilence & 8.76 & 13.30 & 11.51\\
\toprule
\textbf{Violin} & Method &  SDR & SIR & SAR \\
\midrule
               & Spleeter (pretrained)  \cite{Hennequin2019}   & 0.28  & 2.05 & 0.22\\
\hline
                & Random-mixing   & 1.08 & 6.22 & 2.76\\
\hline
               & Wet    & 0.73 & 3.46 & \textbf{3.55}\\ 
               & Chroma & 1.54 & 5.48 & 2.70\\
               & Correlation & \textbf{1.56} & \textbf{8.11} & 2.31\\
               & NonSilence & 1.27 & 6.10 & 2.69\\
\bottomrule

\end{tabular}
\label{tab:result1}
\end{table}

From Tables \ref{tab:result1} and \ref{tab:result2}, we also see that the separation performance generally improves along with the increase in the number of training data, which is not surprising.

\begin{figure}
\centerline{\includegraphics[width = 0.9\columnwidth]{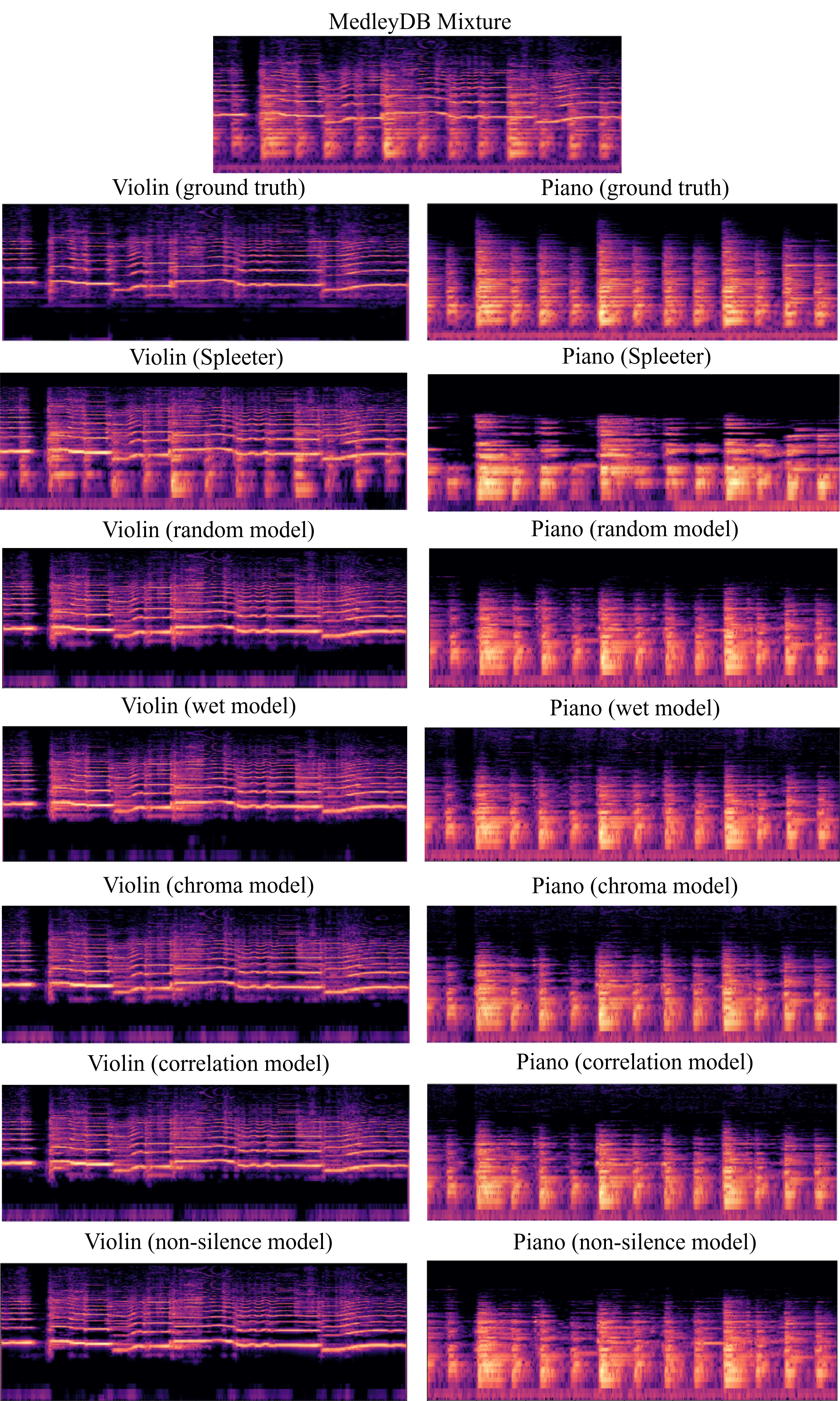}}
\caption{The spectrograms of a mixture from the MedleyDB (titled `MatthewEntwistle\_ImpressionsOfSaturn\_p0\_v1.wav' in our GitHub repo), the ground truth violin/piano stems, and the predicted stems by different methods, trained under the data-rich scenario.}
\label{fig}
\end{figure}

The spectrograms of the original and predicted audio of a text song are shown in Figure \ref{fig}. It can be seen that, the pre-trained Spleeter model cannot separate the piano from the violin well.  In contrast, despite some visible residuals, the models trained with the proposed methods work well in separating the piano and violin. We also see from the result of the violin that some of the proposed methods (e.g., the `wet' model) can get rid of the leakage of the piano part and noises in the low frequency bands, while the baseline methods suffer. 

\begin{figure}
\centerline{\includegraphics[width = 0.9\columnwidth]{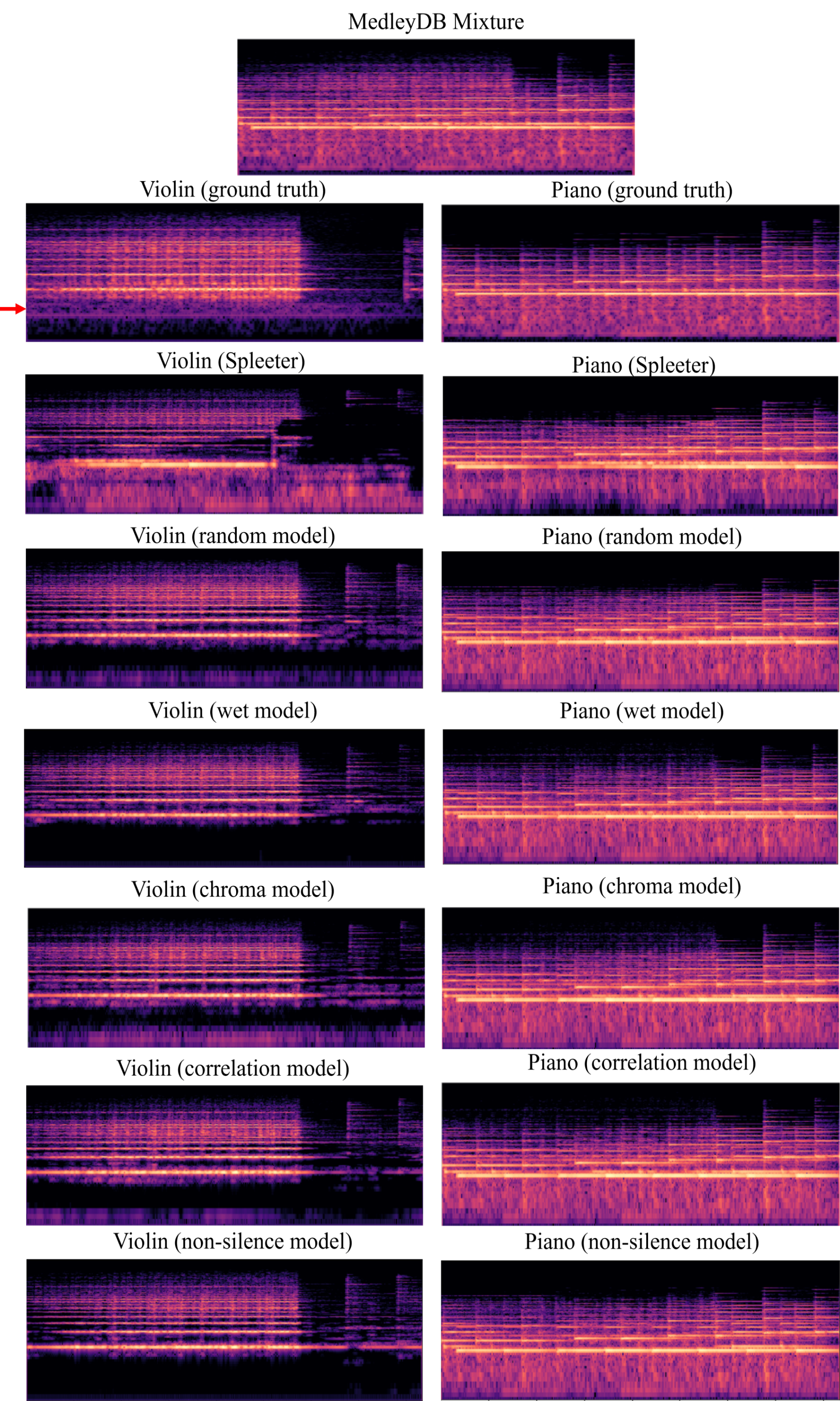}}
\caption{The result of another mixture from the MedleyDB (`MatthewEntwistle\_AnEveningWithOliver\_p0\_v6.wav').}
\label{fig2}
\end{figure}

Figure \ref{fig2} provides another example result. As there is a note offset in the latter half of the violin, the performance difference among the methods can be seen more clearly.
We can see again that the proposed method suffers less from the leakage of the piano in the low frequency bands. 
Besides, the high-frequency piano residuals can be found to be a bit more in the result of the baseline `random' model than in the `wet' model. We also note that an extra low-frequency component (highlighted by the red arrow) in the ground truth violin in this example. Since it is below the lowest frequency of the violin sound (i.e., 196 hz), we listened to it and found that it seems to be some background noise or reverb.
The proposed methods can also exclude this low-frequency part from the separation result.

\begin{table}
\caption{Evaluation result (in db)
for the data-rich scenario} 
\centering
\begin{tabular}{l|lrrr}
\toprule
\textbf{Piano} & Method &  SDR & SIR & SAR \\
\midrule
               & Random-mixing   & \ \textbf{13.46} & 20.76 & \textbf{15.15}\\
\hline
               & Wet    & 11.76 & \textbf{22.38} & 13.76\\ 
               & Chroma & 12.52 & 19.87 & 14.40\\   
               & Correlation & 11.37 & 20.80 & 13.48\\ 
               & NonSilence & 12.82 & 18.85 & 14.98\\

\toprule
\textbf{Violin} & Method &  SDR & SIR & SAR \\
\midrule
                & Random-mixing    & 3.84 & 17.16 & 4.38\\
\hline
               & Wet    & \textbf{4.48} & \ 15.00 & 4.67\\ 
               & Chroma & 3.82 & 16.72 & \textbf{4.76}\\ 
               & Correlation & \ 4.19 & 15.86 & 4.23\\ 
               & NonSilence & 3.03 & \textbf{17.38} & 4.09\\
\bottomrule

\end{tabular}
\label{tab:result2}
\end{table}

\section{Conclusions and Future works}
\label{sec:final}
In this paper, we have proposed and investigated a number of mixing-related data augmentation methods to facilitate the training of deep learning models for music source separation. Result demonstrates the effectiveness of the proposed augmentation methods in the case of small data. When the training data is big enough, the existing random-mixing based approach \cite{Uhlich2017} is strong enough. We have also shown that the implemented models outperform  the current available best model, Spleeter \cite{Hennequin2019}, in violin/piano separation. We believe such training and augmentation methods have potential in benefiting other source separation tasks with limited amount of training data available. In the future, we are interested in extending our experiments to other popular yet less investigated instruments (e.g., guitar and saxophone), and in experimenting with more advanced mixing methods (e.g., applying mixing after the stems are selected, rather than applying mixing to the stems individually beforehand as done here).

\bibliographystyle{IEEEtran}
\bibliography{SourceSep}
\end{document}